\documentclass[a4paper,12pt]{article}
\usepackage{amsfonts}
\usepackage[dvips]{graphics,graphicx}

\newcommand{\be}{\begin{equation}}
\newcommand{\ee}{\end{equation}}
\newcommand{\bea}{\begin{eqnarray}}
\newcommand{\eea}{\end{eqnarray}}
\newcommand{\C}{\mathbb C}
\newcommand{\R}{\mathbb R}
\newcommand{\pa}{\partial}
\newcommand{\im}{\imath}
\renewcommand{\th}{\theta}
\newcommand{\al}{\alpha}
\renewcommand\Im {\mbox{Im }}
\renewcommand\Re {\mbox{Re }}
\newcommand\pr{{\mbox{\scriptsize pure}}}

\begin{document}

\hfill {}
\vskip 0.7 true cm
\begin{center}
{\Large \bf
Differential cross section for Aharonov--Bohm effect with non standard 
boundary conditions}
\end{center}
\vskip 1.5 true cm
\begin{center}
{\large P. \v S\v tov\'{\i}\v cek, O. V\'a\v{n}a}
\end{center}
\begin{center}
{\it Department of Mathematics, Faculty of Nuclear Science, CTU,\\
       Trojanova 13, 120 00 Prague, Czech Republic
}
\end{center}
\vskip 2 true cm
\begin{center}
{\large Abstract}
\end{center}
\bigskip
\noindent
\hspace{.5in}\begin{minipage}{5in}
A basic analysis is provided for the differential
cross section characterizing Aharonov--Bohm effect with
non standard (non regular) boundary conditions imposed on a wave
function at the potential barrier. If compared with the standard
case two new features can occur: a violation of rotational
symmetry and a more significant backward scattering.
\end{minipage}
\vskip 1.5 true cm

\begin{minipage}{4.5in}
\end{minipage}
\vskip 1.5 true cm

\newpage

\noindent
The purpose of this letter is to visualize the results of a recent
paper \cite{1} in which the dynamics of a non-relativistic
spinless quantum particle was  studied 
under the joint effect of a magnetic flux
together with a potential barrier shielding a thin, infinite solenoid.
The new feature of the mathematical model was that it allowed a general
boundary condition imposed on a wave function at the potential barrier. 
Of course, the traditional regular boundary condition, 
as introduced by Aharonov
and Bohm \cite{2}, is included as a particular case.  
We note that the same subject has been  treated independently  in \cite{3}. 
However, the theoretical formulae derived in \cite{1} are
complex enough and don't  provide a direct insight into the character of the
differential cross section so that one is forced to do some elementary
numerical analysis.  Thus our main concern here is to discuss the
scattering problem and to plot a few graphs. Particularly interesting is the 
dependence of the differential cross section on the type of boundary condition, 
and it will be actually shown to be non trivial. 
In addition, we were able to simplify
the  mentioned formulae in some particular cases.

We consider the idealized setup when the radius of the solenoid goes to
zero while the value $\phi$ of the flux of the magnetic field is kept
constant. Moreover, owing to the translational symmetry in the direction
of the solenoid the problem reduces immediately to two dimensions. As
usual, we denote respectively by  $m,\ e$ and $E$ the mass,
the electric charge, and the energy of the scattering particle, and we
set  $k = (2mE/\hbar^2)^{1/2}$.

In \cite{1} a five-parameter family of Hamilton
operators was described. 
One of the parameters is related directly to the flux. Namely, we 
shall use the rescaled quantity 
\be
\label{}
\alpha := -e\phi/2\pi\hbar c\quad\mbox{with}\quad\alpha\in(0,1).
\ee
The restriction of the range of $\alpha$ is possible due to the gauge
symmetry \cite{4}. Actually, as is well known, the quantum particle
cannot
distinguish between two fluxes which differ by an integer multiple of
$2\pi\hbar c/e$. Moreover, we have excluded the
value $\alpha = 0$ corresponding to the vanishing magnetic flux. The
remaining four parameters determine boundary conditions imposed on
the wave function at the origin, and should be related in some way to
the strength and quality of the potential barrier. As already mentioned, 
the usual Aharonov-Bohm (AB) effect \cite{2} corresponds 
to the regular boundary condition, and, for the sake of simplicity, we shall
call it the pure AB effect. 

Let us now describe the family of Hamilton 
operators explicitly. All of them are the usual differential operators 
in the polar coordinates $r,\ \theta$ :
\be
\label{}
-\frac{\hbar^2}{2m}
\left({\pa^2 \over\partial r^2} + {1\over r} ~{\partial\over\partial r} +
{1\over r^2}~\left({\partial\over\partial\theta}+\im\alpha\right)^2 \right).
\ee
To specify the boundary conditions we first introduce the
quantities $\Phi_k^j(\psi)$, $j, k =1, 2$,
describing the asymptotic behavior of a wave function $\psi$ at
the origin: 
\clearpage
\bea
\Phi_{1}^1 (\psi ) &:=& \lim_{r\to 0} ~r^{1-\alpha}
  \int_0^{2\pi} \psi (r, \th )\, e^{\im \th}\, d\th /2\pi \,,
  \nonumber\\
\label{PHIS}
\Phi_{2}^1 (\psi ) &:=& \lim_{r\to 0} ~r^{-1+\alpha}
  \left[ \int_0^{2\pi} \psi (r, \th )\, e^{\im\th}\, d\th /2\pi
            -r^{-1+\alpha}\Phi_{1}^1 (\psi ) \right]\,, \\
\Phi_{1}^2 (\psi ) &:=& \lim_{r\to 0} ~r^{\alpha}
  \int_0^{2\pi} \psi (r, \th )\, d\th /2\pi \,,
  \nonumber\\
\Phi_{2}^2 (\psi ) &:=& \lim_{r\to 0} ~r^{-\alpha}
  \left[ \int_0^{2\pi} \psi (r, \th )\, d\th /2\pi
            -r^{-\alpha}\Phi_{1}^2 (\psi ) \right] \,.\nonumber
\eea
The boundary conditions then read
\be
\label{BC}
\pmatrix{
\Phi_{1}^1 (\psi )\cr\Phi_{1}^2 (\psi )}
=
\pmatrix{  u' & \alpha\bar{w'} \cr
        (1\! -\!\alpha )w' & v' }
\pmatrix{
\Phi_{2}^1 (\psi )\cr
\Phi_{2}^2 (\psi )}
\ee
where $u',v'\in\R$ and $w'\in\C$ represent altogether four real parameters. 
Particularly 
the pure AB effect corresponds to the values $u'=v'=0$ and $w'=0$. 
We note also that the boundary conditions (4) are rotationally invariant 
only if $w' = 0$. So generally the angular momentum is not conserved.

To make simpler the formulae presented below we will use the dimensionless 
parameters
\be
u := \frac{\Gamma(\alpha)}
{\Gamma(2-\alpha)}\left({k\over2}\right)^{2-2\alpha}\,u',\ 
v := \frac{\Gamma(1-\alpha)}
{\Gamma(1+\alpha)}\left({k\over2}\right)^{2\alpha}\,v',\ 
w := {k\over2}\,w'.
\ee
But one has to keep in mind that $u,\ v$ and $w$  now depend on the momentum 
$k$ and that the true  parameters fixing the Hamilton operator are the original
ones, i.e., $u',\ v'$ and $w'$.

Let us recall that the differential cross section in the plane is given by 
the equality
\be
\frac{d\sigma(\theta)}{d\theta}  =
{2\pi\over k}\vert S(k;\theta ,\theta_0)\vert^2 
\ee
where $S(k;\theta ,\theta_0)$ is the scattering matrix. The angle 
$\theta_0$ determines the direction of motion of 
the incident particle and it is generally of importance because of the 
violation of rotational  symmetry when $w\neq 0$. In fact, if the problem 
was rotationally  symmetric the angles $\theta$ and $\theta_0$ would occur 
in the expression for 
$S(k;\theta ,\theta_0)$ only in the combination  $\theta-\theta_0$ which need
not be  the case as we shall see. Thus one has to consider the dependence of
the differential  cross section altogether on six real parameters: 
$u,\ v,\ \Re w,\ \Im w,\ \alpha$ and  $\theta_0$. 

For the scattering matrix the following formula has been derived:  
\bea
S(k;\theta , \theta_0 ) &=&
\cos(\pi\alpha)\,\delta (\theta\! -\!\theta_0 ) + {1\over 2\pi} \sin(\pi\alpha) 
\frac{ e^{-\im(\theta - \theta_0)/2} }{\sin\bigl((\theta - \theta_0) /2\bigr)}
 \\
\label{SM}
&&+ {1\over 2\pi} \left( 
(\Sigma_{11} - e^{\im\pi\alpha}) e^{-\im(\theta - \theta_0)}+
\Sigma_{12}\, e^{-\im\theta}+\Sigma_{21}\, e^{\im\theta_0}+
\Sigma_{22} - e^{-\im\pi\alpha}  \right) \nonumber
\eea
where 
\bea
\Sigma_{11} &=& {\det}^{-1}N(k)\,
\left(e^{-\im\pi\alpha} \left(uv - |w|^2\right) +
u -v  -e^{\im\pi\alpha}  \right) , \nonumber \\
\Sigma_{12}
&=& - {\det}^{-1}N(k)\, 2\im \,\sin(\pi\alpha )~\bar w , \nonumber\\
\label{SIGMAS}
\Sigma_{21} &=& - {\det}^{-1}N(k)\, 2\im \,\sin(\pi\alpha )~ w , \\
\Sigma_{22} &=& {\det}^{-1}N(k)\, 
\left(e^{\im\pi\alpha} \left(uv - |w|^2\right) +
u -v  -e^{-\im\pi\alpha}  \right) , \nonumber 
\eea
and
\be
\label{DETN}
\det N(k) = uv - |w|^2 +  e^{\im\pi\alpha}\, u - e^{-\im\pi\alpha}\, v -1.
\ee
Concerning the differential cross section, 
after some manipulations we arrive at the expression 
\bea
\frac{d\sigma(\theta)}{d\theta}  = \frac{2\sin^2(\pi\alpha)}{\pi k} 
\left| 
 \frac{1}{2\sin\bigl((\theta - \theta_0) /2\bigr)}-
\frac{1}{uv - |w|^2 +  e^{\im\pi\alpha}\, u - e^{-\im\pi\alpha}\, v -1}
\right.\nonumber \\
\times \left( 
2\sin\bigl((\theta - \theta_0) /2\bigr) (uv - |w|^2) + 
\im\, e^{\im\bigr(\pi\alpha-(\theta-\theta_0)/2\bigl)}\, u  
\right.\\ 
\left.\left.
+\im\, e^{-\im\bigr(\pi\alpha-(\theta-\theta_0)/2\bigl)}\, v + 
2\im\,\mbox{Re}\biggl(e^{\im(\theta + \theta_0)/2}\, w\biggr)
\right) \right|^2 . \nonumber
\eea


Let us proceed to the discussion of the behavior of the function 
$2\pi\vert S(k;\theta ,\theta_0)\vert^2 $. For the sake of convenience, 
in the graphs presented below this function
depends on the angle $\Theta = \theta - \theta_0 + \pi$
(mod $2\pi$) rather than directly on $\theta$. Hence the values 
$\Theta = 0$ and  $\Theta = -\pi$ correspond to the backward and forward
scattering, respectively. 

\begin{figure}[h]
%
%
\includegraphics[width=0.90\linewidth]{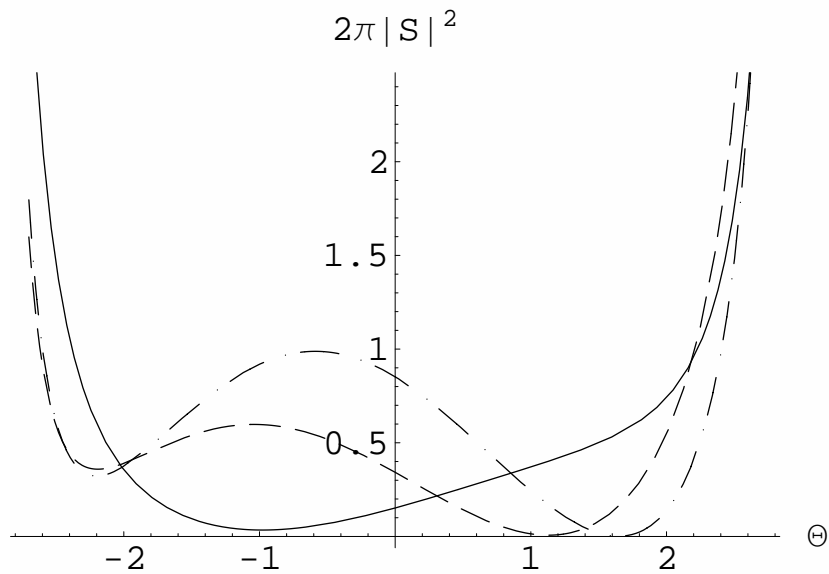}
\caption{Dependance of $2\pi |S( k;\theta, \theta_0)|^2$ 
on $\Theta = \theta - \theta_0 + \pi$ (mod $2\pi$), the solid line corresponds to 
$u = 25$, $v = 1$, $w = 3 + 3 \im$ , the dashed line corresponds to 
$u = 20$, $v = 0$, $w = 3$, the dot-dashed line corresponds to 
$u = 1$, $v = 10$, $w = 0$, and  $\alpha = 0.5$, $\theta_0 =  0$ 
in all three cases.}
\label{fig1}
\end{figure}

On Figure~\ref{fig1} we show graphs for three
different boundary conditions which have been chosen rather accidentally, 
and in all three cases $\alpha = 0.5$ and $\theta_0 = 0$. 
As one can observe, there is
a common feature which is independent of the boundary conditions and of the
angle $\theta_0$. The differential cross section is divergent for 
$\Theta$ tending to  $\pm\pi$ (forward scattering).  The explanation is
simple. The considered problem is somewhat inconsistent from the physical point
of view as the total magnetic flux passing through the plane is nonzero. A
more consistent arrangement  would involve two parallel solenoids with equal
fluxes but oppositely oriented  \cite{5}. In this case the divergence is 
actually removed, as discussed in \cite{6}. 

Further, a rough inspection of the graphs leads
to the conclusion that there are two possible shapes. Either the graph
exhibits one minimum as in the pure AB effect or there are two local minima
and one local maximum. The latter shape takes place for 
some non-standard boundary conditions and implies existence of a small and
rather flat peak centered closely at the value $\Theta = 0$. This suggests 
that, at least in principle, one should be able to detect the boundary
conditions describing the physical situation when looking at the backward
scattering picture.

To illustrate this observation let us now consider more closely the particular
case with $u = v = 0$. Then the formulae simplify significantly. 
It is convenient to write $w$ in the polar form, $w = \rho\exp(\im\varphi)$. 
The differential cross section then reads 
\bea
\frac{d\sigma(\theta)}{d\theta} &=& 
\frac{\sin^2(\pi\alpha)}{2\pi k} \\
& & \times\left( 
 \frac{1}{\sin^2\bigl((\theta - \theta_0) /2\bigr)} +
8\frac{\rho^2}{(1+\rho^2)^2}
\left( 
\cos(\theta + \theta_0 +2\varphi) - 
\cos(\theta - \theta_0)\:\rho^2
\right) \right) . \nonumber
\eea
The value $w = 0$ corresponds to the pure AB effect, and then 
\be
\frac{d\sigma_\pr (\theta)}{d\theta} = 
\frac{\sin^2(\pi\alpha)}{2\pi k\sin^2\bigl((\theta - \theta_0) /2\bigr)}
\,. 
\ee

Let us note that though the differential cross section
diverges and so the total cross section is not well defined one can take the
pure AB effect for the reference point and integrate the difference of the
differential cross sections. The result is obviously 
\be
\int_0^{2\pi}\left(\frac{d\sigma(\theta)}{d\theta}-
\frac{d\sigma_\pr (\theta)}{d\theta}\right)\,d\theta = 0.
\ee 

As one can see from (11),
the magnetic flux enters the formula in the form of a prefactor 
$\sin^2(\pi\alpha)$. The dependence on the initial angle 
$\theta_0$ as well as on the argument $\varphi$ of $w$ is rather
weak. However there is a remarkable difference in the shape of the graph for 
$w = 0$  and $|w|$ large. Actually it is not difficult to calculate the limit 
of the differential cross section for $|w|\to\infty$ (with $u = v = 0$). 
This way we get the formula 
\be
\frac{d\sigma_{|w|\to\infty}(\theta)}{d\theta} = 
\frac{\sin^2(\pi\alpha)\left(1-2\cos(\theta - \theta_0)\right)^2}
{2\pi k\sin^2\bigl((\theta - \theta_0) /2\bigr)} \,. 
\ee

\begin{figure}[h]
\includegraphics[width=0.90\linewidth]{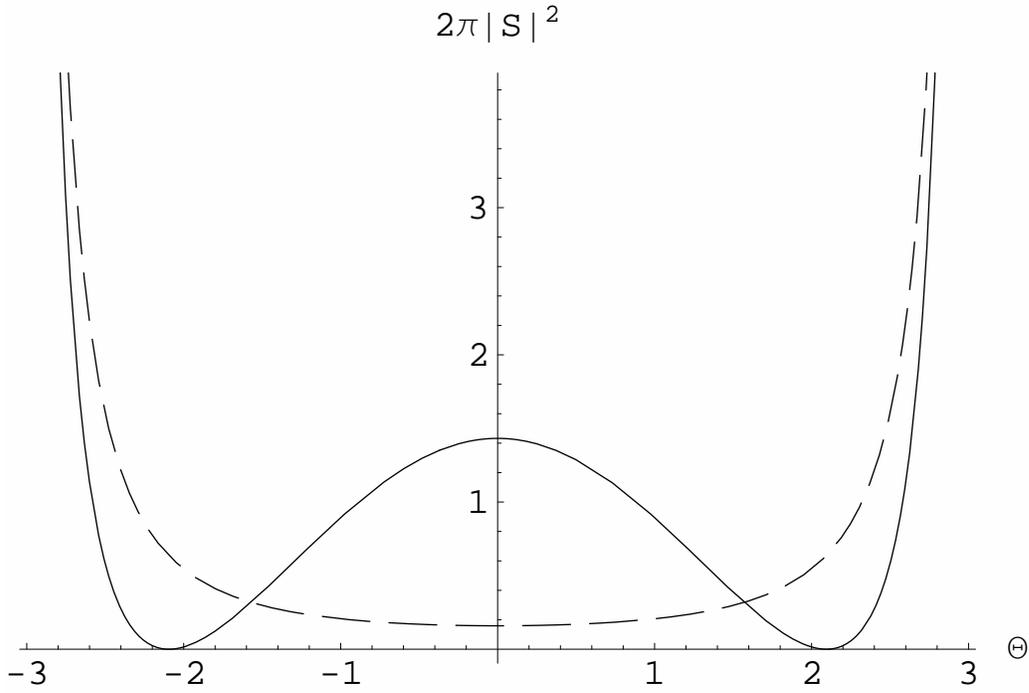}
\caption{Dependance of $2\pi |S( k;\theta, \theta_0)|^2$ 
on $\Theta = \theta - \theta_0 + \pi$ (mod $2\pi$), the solid line corresponds 
to (14) , the dashed line corresponds to (12), $\alpha = 0.5$, and 
$\theta_0$ can be arbitrary.}
\label{fig2}
\end{figure}

This limit procedure can be interpreted in two ways. Either one
assumes that $u' = v' = 0$,  $w'$ is fixed and the energy of the particle 
is large, or that the energy is constant while  
$|w'|\to\infty$  (c.f. (5)). As one
finds immediately from (4), the latter interpretation corresponds to the
boundary conditions 
\be
\Phi_2^1(\psi) = \Phi_2^2(\psi) = 0.
\ee
Let us compare the formula (14) with the analogous formula (12) for
the pure AB effect. Figure~\ref{fig2} depicts the two graphs. 

\begin{figure}[h]
\includegraphics[width=1.2\linewidth]{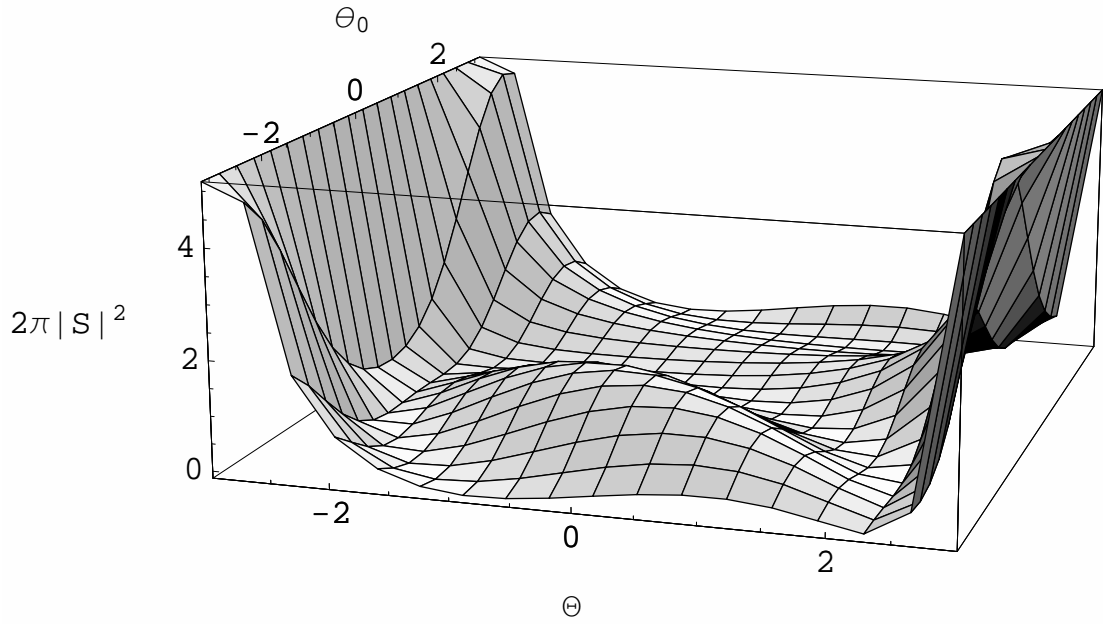}
\caption{Dependance of $2\pi |S( k;\theta, \theta_0)|^2$ 
on $\Theta = \theta - \theta_0 + \pi$ (mod $2\pi$) and $\theta_0$ for
$u=v=w=5$, , $\alpha = 0.5$.}
\label{fig3}
\end{figure}

Let us now examine another particular case, this time with  
$u = v = w > 0$, hence  $uv - |w|^2 = 0$. Then we have 
\bea
\frac{d\sigma(\theta)}{d\theta} &=& 
\frac{\sin^2(\pi\alpha)}
{2\pi k\sin^2\bigl((\theta - \theta_0) /2\bigr)}\\
&& \times \frac{1+4\big(\sin\theta-\sin\theta_0-
                \sin(\pi\alpha-\theta+\theta_0)\big)^2\,u^2}
{1+4\sin^2(\pi\alpha)\:u^2}\nonumber
\,. 
\eea
Here we can demonstrate a clear violation of the rotational symmetry. Indeed, 
the three-dimensional plot given in Figure~\ref{fig3} illustrates its rather strong 
dependence on the initial angle $\theta_0$. 

This case also indicates that the equality (13) need not be true in general. 
Comparing again the differential cross 
section (16) to that one related to the pure AB effect we obtain 
\be
\frac{d\sigma(\theta)}{d\theta}-\frac{d\sigma_\pr(\theta)}{d\theta}= 
\frac{\sin^2(\pi\alpha)\,u^2}
{\pi k(1 + 4\sin^2(\pi\alpha)\,u^2)}
\left(f(\al,\th,\th_0) -\frac{\cos^3\bigl((\theta - \theta_0) /2\big)}
{\sin\bigl((\theta - \theta_0) /2\bigr)}
\right)
\ee
where
\bea
f(\al,\th,\th_0) 
&=& 8\cos(\pi\alpha)\: (\cos(\pi\alpha)+\cos\theta_0) +
7\cos(\pi\alpha-\theta) + \cos(\pi\alpha+\theta)\nonumber\\
&& - 2\sin(\pi\alpha)\: \sin(\theta-2\theta_0) + 
\cos(2\pi\alpha+\theta-\theta_0)  \\
&& + 3\cos(2\pi\alpha-\theta+\theta_0) + 
4\cos(\theta+\theta_0) \,. \nonumber
\eea
Even this expression still contains a nonintegrable singularity, namely the 
term $\cos^3\bigl((\theta - \theta_0) /2\bigr)/
\sin\bigl((\theta - \theta_0) /2\bigr)$. 
However since this function 
is $2\pi$ periodic and odd with respect to the point $\theta_0$ 
we can set its integral over an interval of length 
$2\pi$  equal 0. With this assumption we find that 
\be
\int_0^{2\pi}\left(
\frac{d\sigma(\theta)}{d\theta} - \frac{d\sigma_\pr(\theta)}{d\theta}
\right)\, d\theta = 
\frac{16\sin^2(\pi\alpha)\cos(\pi\alpha)
\big(\cos(\pi\alpha) + \cos\theta_0\big)\,u^2}
{ k(1 + 4\sin^2(\pi\alpha)\,u^2)}
\ee
which generally need not vanish. 

Finally let us consider the case with conserved angular momentum which means 
that $w = 0$.  
Then the differential cross section equals 
\be
\frac{d\sigma(\theta)}{d\theta} = 
\frac{\sin^2(\pi\alpha)\: g(u,v,\th,\th_0)}
{2\pi k\sin^2\bigl((\theta - \theta_0) /2\bigr)
\bigl(1+u^2-2u\cos(\pi\alpha)\bigr)
\bigl(1+v^2+2v\cos(\pi\alpha)\bigr)}
\ee
where 
\bea
g(u,v,\th,\th_0) &=& 
(1+u^2)(1+v^2)+4uv\sin^2(\pi\alpha-\theta+\theta_0)+
4u^2v^2\cos^2(\theta-\theta_0) \nonumber\\
&& - 2(u-v)(1+uv)\cos(\pi\alpha-\theta+\theta_0)-
4uv(1+uv)\cos(\theta - \theta_0) \nonumber\\
&& +4(u-v)uv\cos(\pi\alpha-\theta+\theta_0)
\cos(\theta - \theta_0) \,. 
\eea
Specializing even more, namely setting $u = v$ and $\alpha = 1/2$, 
we get
\be
\frac{d\sigma(\theta)}{d\theta} = 
\frac{1+u^2\left(1-2\cos(\theta - \theta_0)\right)^2}
{2\pi k\left( 1+u^2 \right)\,
\sin^2\bigl((\theta - \theta_0) /2\bigr)} 
\ee   
This expression quite resembles the case with $u = v = 0$, 
$w \neq 0$, particularly the limit procedure $u\to\infty$  
leads again to the formula (14) (but with $\alpha = 1/2$).

This concludes our brief analysis of the differential cross
section in dependence on the choice of parameters characterizing the nature of
the potential barrier. In fact, it would be not difficult for anyone interested
in to reexamine or prolong this analysis when starting from the formula (10).
Basically we have demonstrated two new features which may occur: a more
significant backward scattering (c.f. Fig.~\ref{fig2}) and a violation of rotational
symmetry (c.f. Fig.~\ref{fig3}). 


\vskip 24pt
{\bf Acknowledgements.}\
P.S. wishes to  gratefully acknowledge the partial
support from Grant No. 202/96/0218 of Czech GA.

\newpage
\section*{Figures}
%

\newpage

\newpage


\end{document}